\documentclass[a4paper,11pt]{article}

\usepackage{german}
\usepackage{times}
\usepackage{supertabular}

\usepackage{amssymb}

\usepackage{graphicx}

\begin{document}

\title{Moore and more and symmetry%
\author{Tobias Kretz and Michael Schreckenberg %
\\
\\
Physik von Transport und Verkehr\\ Universit\"{a}t Duisburg-Essen\\ D-47057 Duisburg, Germany\\ 
}
}

\maketitle

\begin{abstract}
In any spatially discrete model of pedestrian motion which uses a regular lattice as basis, there is the question of how the symmetry 
between the different directions of motion can be restored as far as possible but with limited computational effort. This question is equivalent 
to the question {\it ''How important is the orientation of the axis of discretization for the result of the simulation?''} An optimization in terms of symmetry can be combined with the implementation of higher and heterogeniously distributed walking speeds by representing different walking speeds via different amounts of 
cells an agent may move during one round. Therefore all different 
possible neighborhoods for speeds up to $v=10$ (cells per round) will be examined for the amount of deviation from radial symmetry. Simple criteria will be stated which will allow to find an optimal 
neighborhood for each speed. It will be shown that following these criteria even the best mixture of steps in {\it Moore and von Neumann 
neighborhoods} is unable to reproduce the optimal neighborhood for a speed as low as 4.
\end{abstract}
\section{Introduction}
\begin{figure}[htbp]
\begin{center}
\includegraphics[height=100pt]{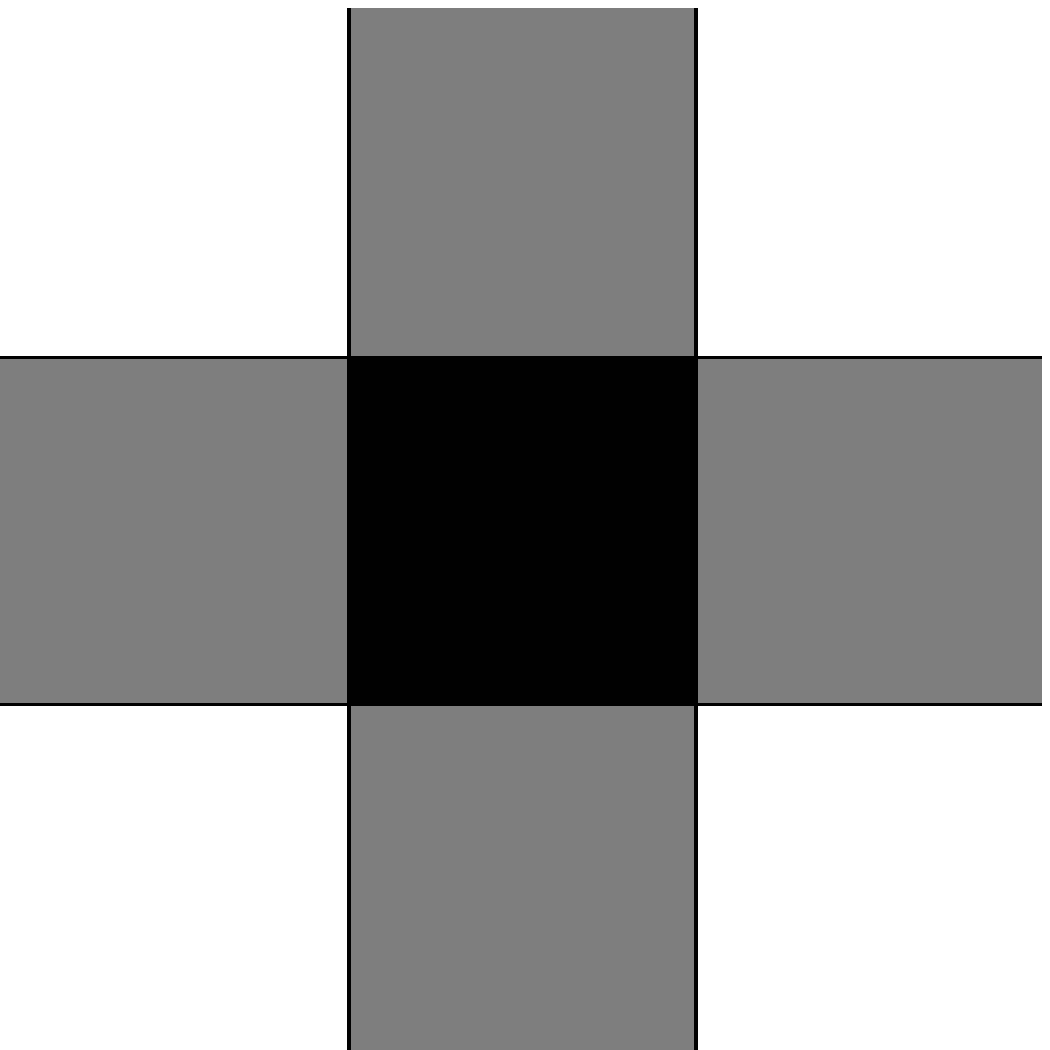}\hspace{12pt}
\includegraphics[height=100pt]{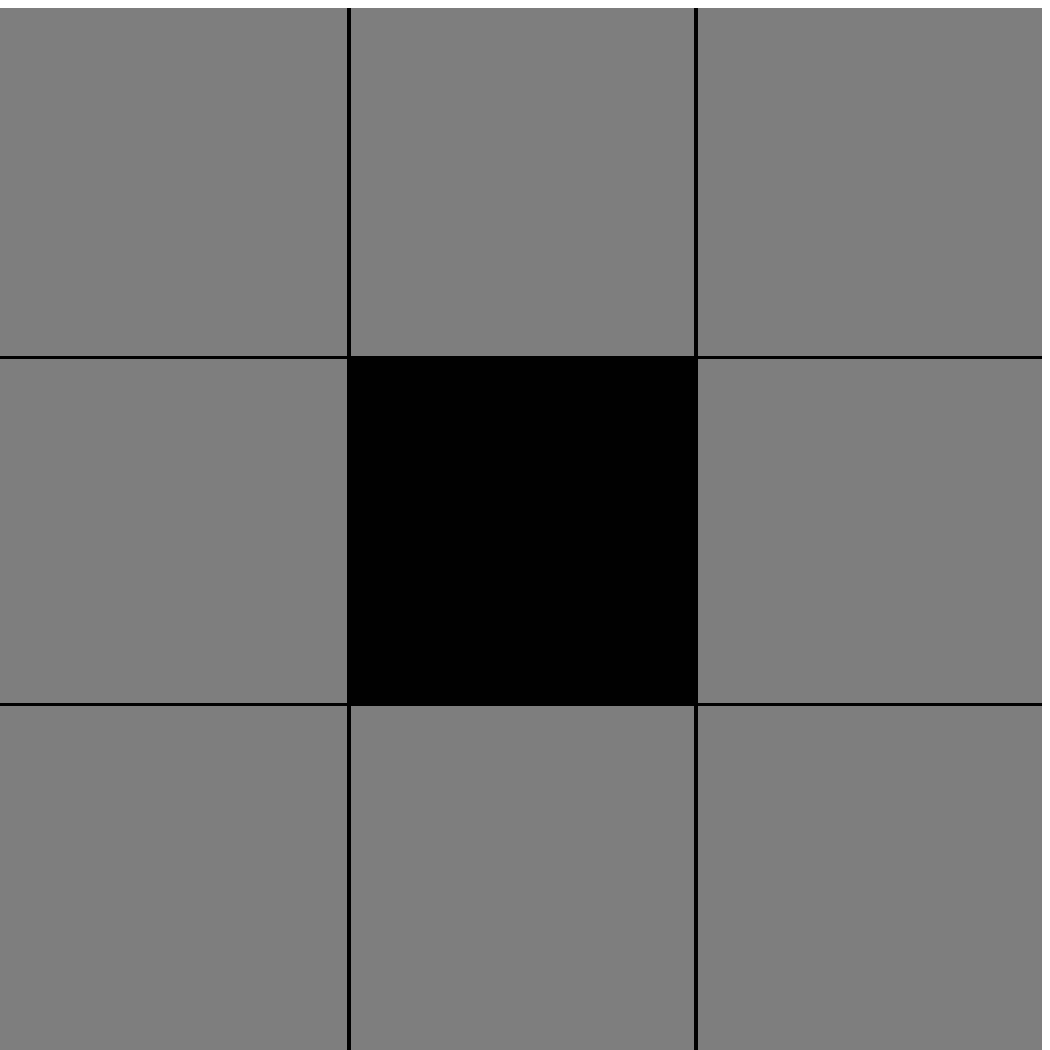}
\caption{\it von Neumann- and Moore-neighborhood (v=1)}
\label{fig:umgebung1}
\end{center}
\end{figure}
\begin{figure}[htbp]
\begin{center}
\includegraphics[height=100pt]{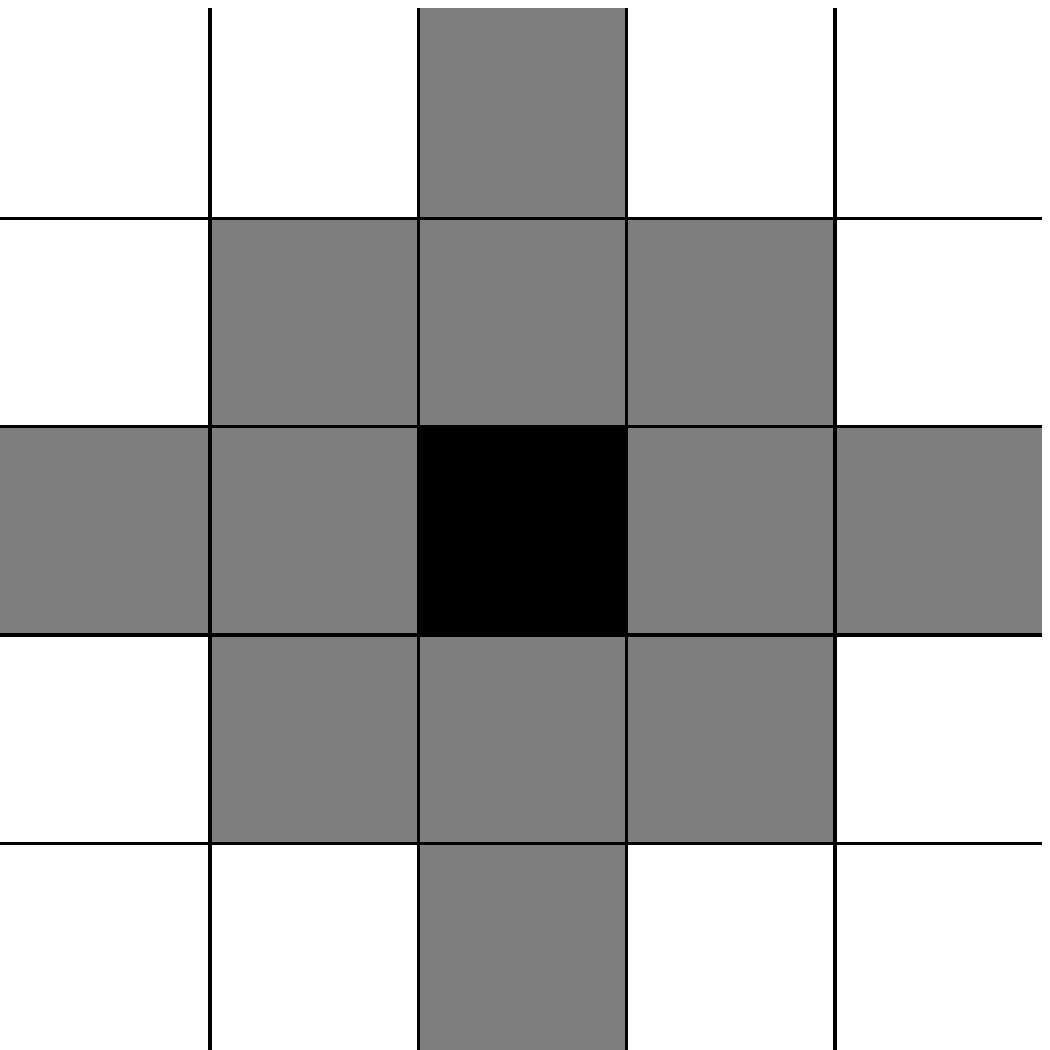}\hspace{12pt}
\includegraphics[height=100pt]{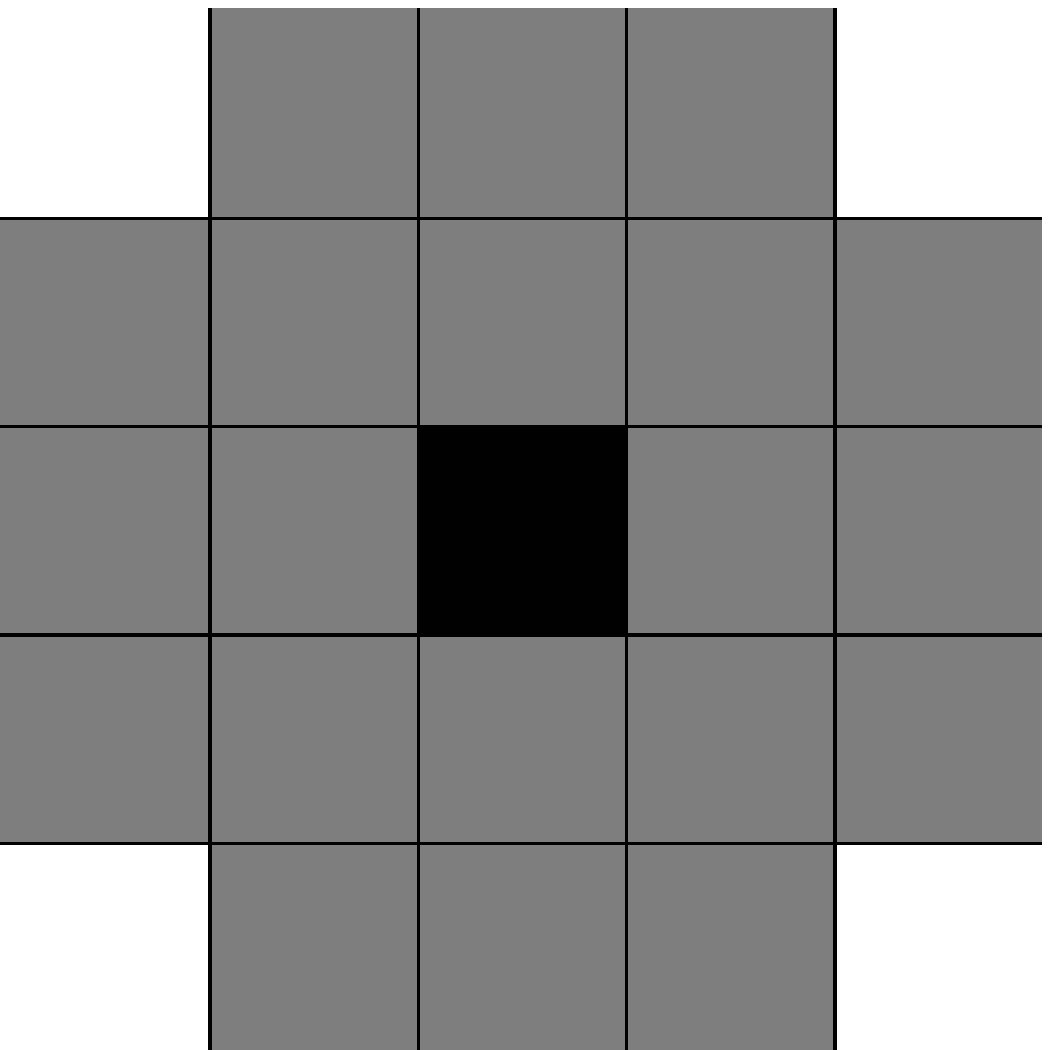}\hspace{12pt}
\includegraphics[height=100pt]{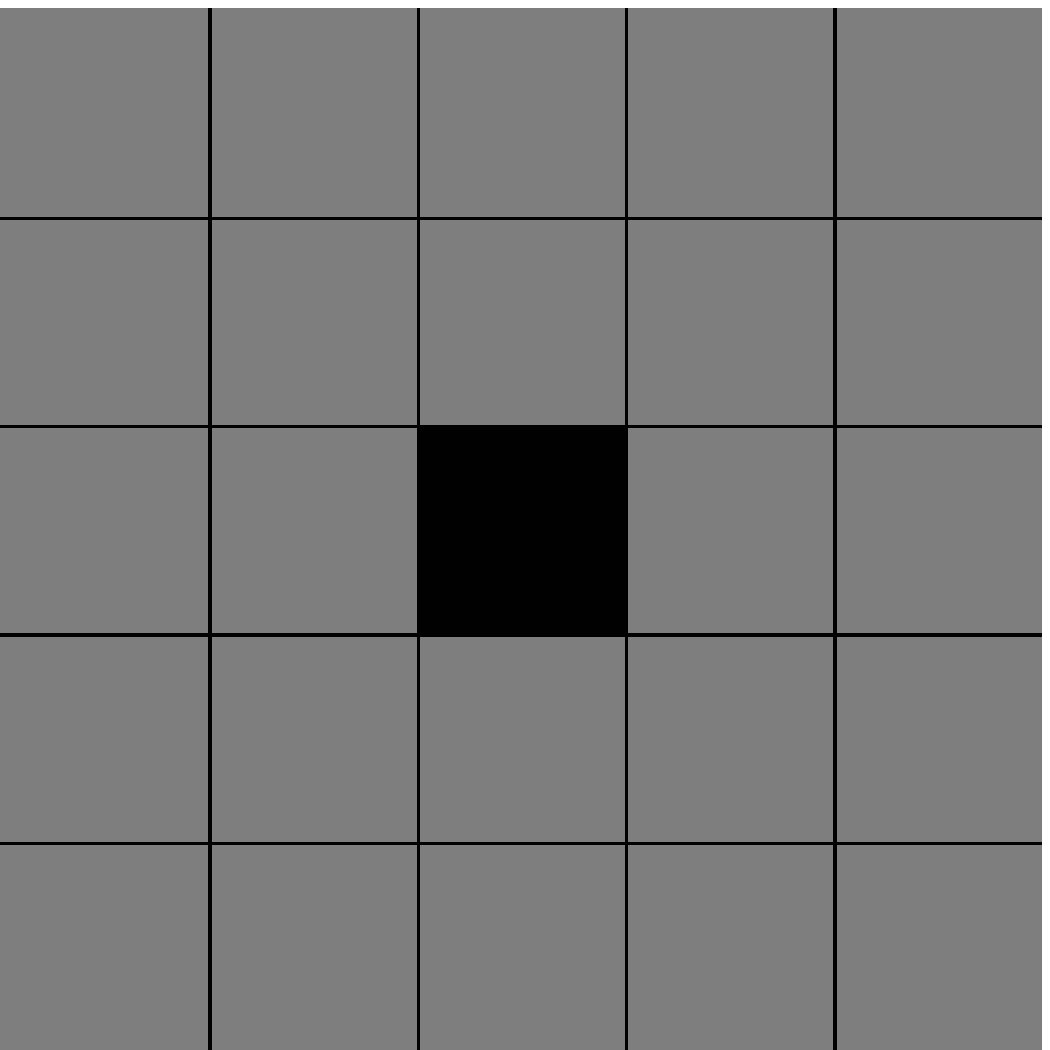}\hspace{12pt}
\caption{\it Complete neighborhoods for v=2}
\label{fig:umgebung2}
\end{center}
\end{figure}
\begin{figure}[htbp]
\begin{center}
\includegraphics[height=100pt]{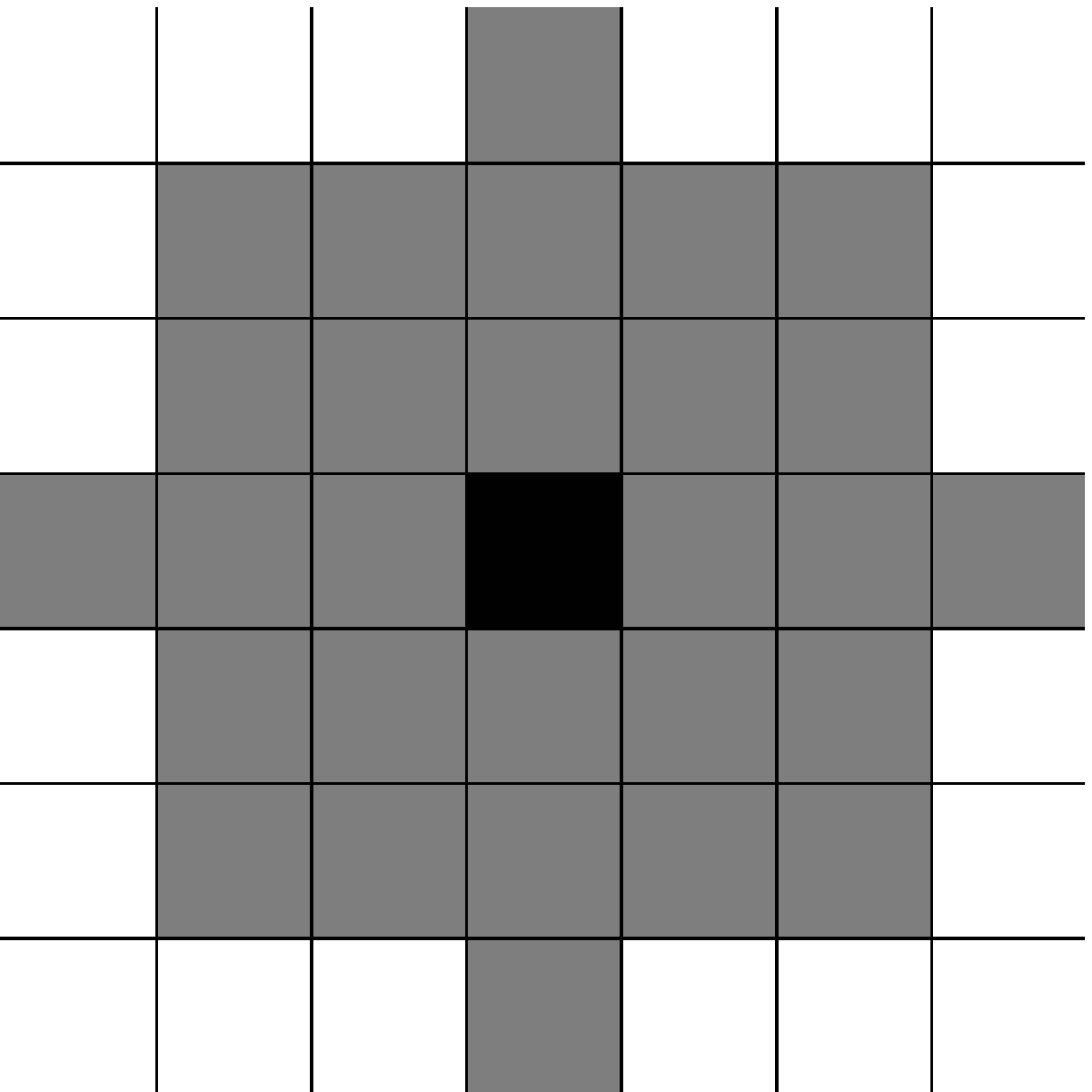}\hspace{12pt}
\includegraphics[height=100pt]{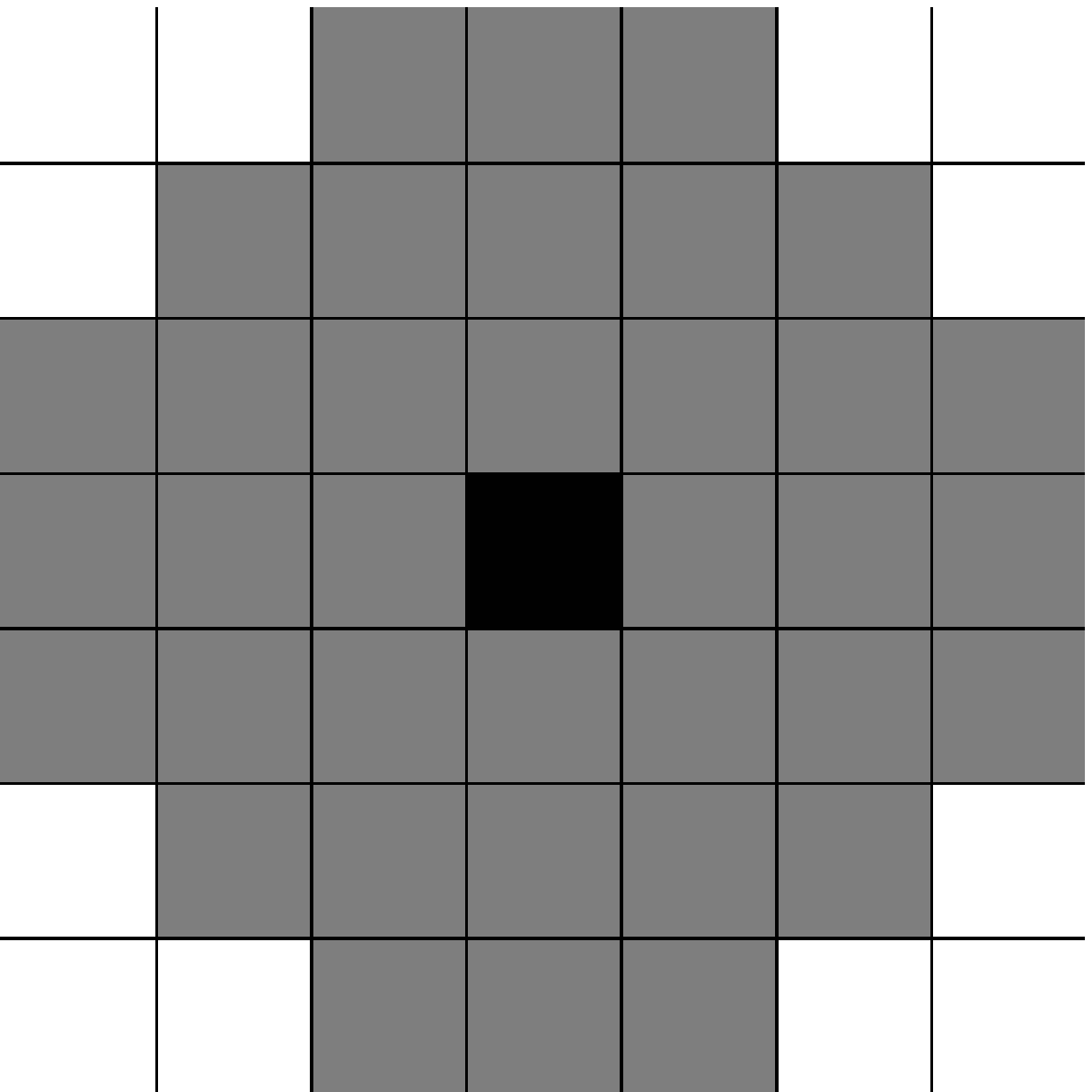}\hspace{12pt}
\includegraphics[height=100pt]{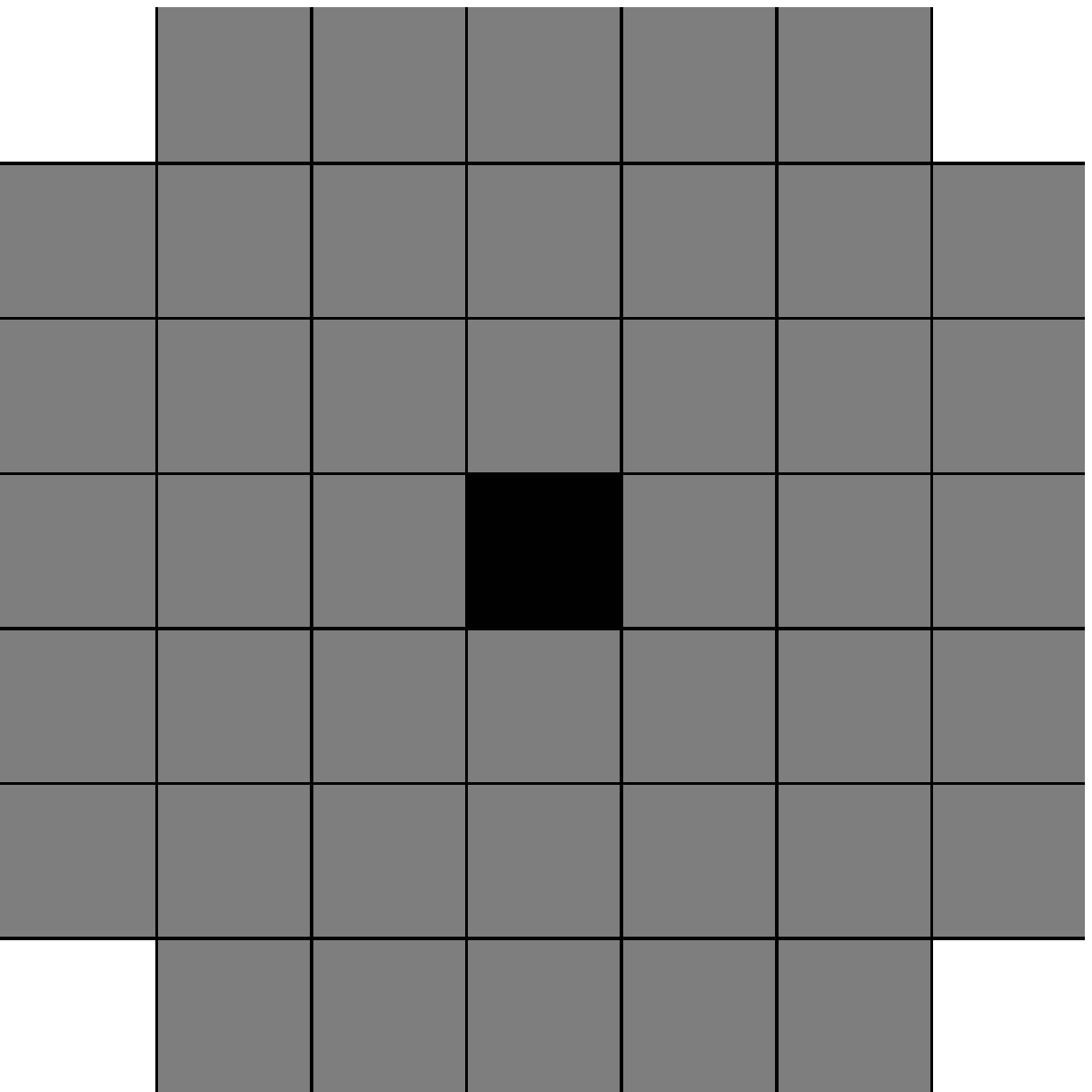}\hspace{12pt}
\caption{\it Complete neighborhoods for v=3}
\label{fig:umgebung3}
\end{center}
\end{figure}

In a model which is spatially and temporally discrete the speed of an agent (as the model of a person in a simulation will be called) is the number of cells which he moves  during one round. As the real-world interpretation of the size of a cell is fixed by the scale of the discretization, the real-time interpretation of one round fixes the real-world interpretation of such a dimensionless speed.\\
Subsequent steps within one of the neighborhoods of Fig. \ref{fig:umgebung1}, leave an agent to be either $\sqrt{2}$ times as slow or as fast moving into the diagonal direction than moving horizontally or vertically. To some extent the situation can be improved by a {\it mixture} of von Neumann and Moore neighborhoods. If for example an agent would be allowed to do a total of five steps during one round, of which three are in von Neumann and two in  Moore neighborhood, a larger {\it total neighborhood} of cells which can be reached during one round would result. The question is: Is there an {\it optimal total neighborhood} for a given speed? And can it be composed of von Neumann and Moore neighborhoods? 
In vertical and horizontal direction there is no doubt about the neighborhood: For a speed $v=v_m$ the neighborhood contains the cell of 
the agent and $v_m$ cells in each horizontal or vertical direction. For any other direction there are cells for which  
it is not obvious if they should be part of the neighborhood. At the very beginning for $v=1$ there is the question if one should use the
von Neumann or Moore neigborhood. (See Fig. \ref{fig:umgebung1}).\\
{\bf Definition:} \em Complete neighborhoods \em are four-fold symmetrical neighborhoods where all cells which belong to the neighborhood 
are closer to the center cell than those which do not.\\
Example: There are three complete neighborhoods for each $v=2$ and $v=3$. (See Figs. \ref{fig:umgebung2}
and \ref{fig:umgebung3}.)\\
\\
\\
Obviously one can limit the search for an optimal neighborhood to complete neighborhoods.\\
The question is, which complete neighborhood represents the corresponding speed (which is number of cells in horizontal and vertical direction) best. However there might exist other alternatives, the criteria
chosen here are such that discretization effects concerning the axis of discretization of the original plan are minimized.
Therefore at first for each complete neighborhood the speed $v(\phi)$ into each direction has to be written down.\\
Then the direction-averaged speed is calculated:
\begin{equation}
<v>=v_{av}=\frac{1}{2\pi}\int_{\phi} v(\phi) d\phi
\end{equation} 
After that the squared deviation of speeds into each direction from this average is calculated.
\begin{equation}
\Delta v=\sqrt{\frac{1}{2\pi}\int_{\phi} (v(\phi)-v_{av})^2 d\phi}
\end{equation}
The criteria for an optimal neighborhood are then
\begin{itemize}
\item The direction averaged speed should be close to the corresponding integer.
\item The deviation from this average into different directions should be small.
\end{itemize}
\paragraph{Complete neighborhoods up to $v=10$:}
The neighborhoods are named by the maximum squared distance of a cell from the center, implying that neighborhood $Y$ includes all cells of neighborhood
$X \leq Y$. For reasons of clarity in the following table the numbers are only shown in the second octant.
\begin{center}
\begin{tabular}{|c|c|c|c|c|c|c|c|}\hline
100&101&104&109&116&   &   &   \\ \hline
 81& 82& 85& 90& 97&106&117&   \\ \hline
 64& 65& 68& 73& 80& 89&100&113\\ \hline
 49& 50& 53& 58& 68& 74& 85&98 \\ \hline
 36& 37& 40& 45& 52& 61& 72&   \\ \hline
 25& 26& 29& 34& 41& 50&   &   \\ \hline
 16& 17& 20& 25& 32&   &   &   \\ \hline
  9& 10& 13& 18&   &   &   &   \\ \hline
  4&  5&  8&   &   &   &   &   \\ \hline
  1&  2&   &   &   &   &   &   \\ \hline
{\bf 0}&   &   &   &   &   &   &   \\ \hline
\end{tabular}\\
\end{center}
{\it NB: Of neighborhoods in the previous table only the neighborhoods 1, 2, 4, 5, 8, 10, 13, 17, 20, 29, 34, 40, 45, 58, 80 and 97 can be composed of the corresponding number of 
subsequent steps within von Neumann and Moore neighborhoods. (Take $N$ von Neumann and $M=v-N$ Moore steps and check how the largest possible neighborhood looks like.)}\\

\subsection{{\it v($\phi$)} - the variation of speed with the direction of motion}
\begin{figure}[htbp]
\begin{center}
\includegraphics[height=200pt]{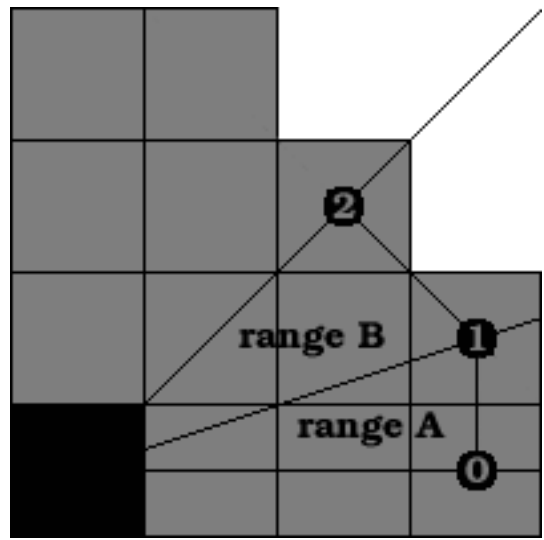}\hspace{12pt}
\caption{\it Example for calculating $v(\phi)$ for one of the v=3 neighborhoods}
\label{fig:umgebung4}
\end{center}
\end{figure}

\begin{figure}[htbp]
\begin{center}
\includegraphics[height=250pt]{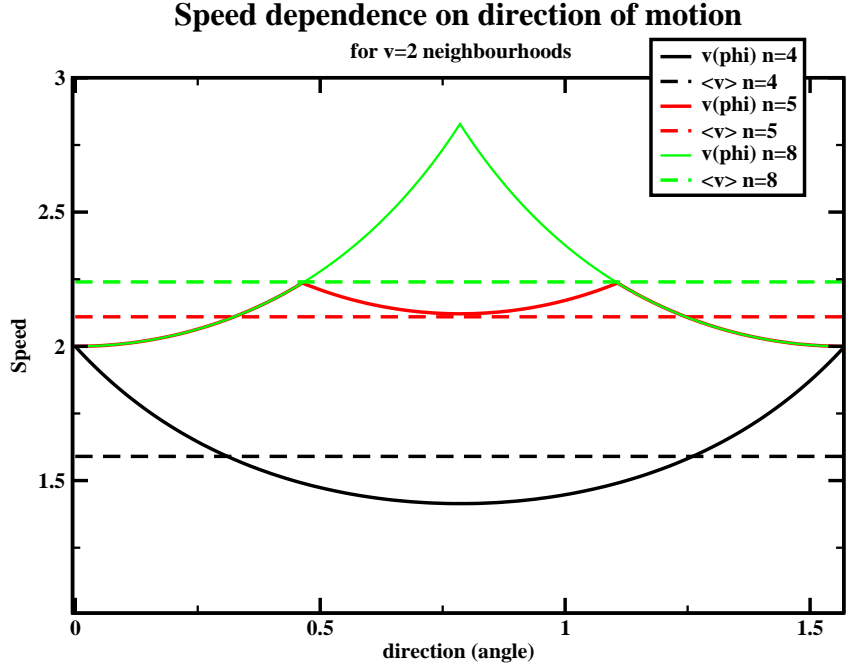}\hspace{12pt}
\caption{\it Example for the $v(\phi)$ dependence. ($v=2$ neighborhoods 4, 5 and 8)}
\label{fig:vphi}
\end{center}
\end{figure}

Since all complete neighborhoods have a fourfold axe-symmetry it is sufficient to calculate $v(\phi)$
for $0\leq \phi < \pi/4$.\\
$v(\phi)$ is continuously composed from different functions resulting from different ranges of $\phi$.
The structure of those ranges depends on the shape of the edge of the neighborhood (See Fig. \ref{fig:umgebung4}).
\\
\vspace{12pt}
Definitions:\\
\begin{tabular}{ c   l}
$\delta x_i$& horizontal distance (in cells) of a border cell to the origin (the\\
            & black cell in Fig. \ref{fig:umgebung4}); with $i$ starting with 0 at $\phi=0$ \\
            & (see the numbering in the black circles in Fig. \ref{fig:umgebung4})\\
$\delta y_i$& vertical distance of a border cell to the origin; with $i$ starting\\
            & with 0 at $\phi=0$\\
$D$         & a large distance\\
$\Delta X$  &$D \cos(\phi)$\\
$\Delta Y$  &$D \sin(\phi)$\\
$N$         & total (minimal) number of steps to reach a cell in distance $D$\\
            & into direction $\arctan(\frac{\delta y_i}{\delta x_i})\leq \phi_i < \arctan(\frac{\delta y_{i+1}}{\delta x_{i+1}})$\\
$n$         &number of steps into direction $(0/0) \rightarrow (\delta x_i / \delta y_i )$\\
$N-n$       &number of steps into direction $(0/0) \rightarrow (\delta x_{i+1} / \delta y_{i+1} )$\\
\end{tabular}\\
To reach the point $(\Delta X /\Delta Y)$ an agent has to do $n$ times a $(\delta x_i / \delta y_i )$ step 
and $(N-n)$ times a $(\delta x_{i+1} / \delta y_{i+1} )$ step.\\
Such that
\begin{eqnarray}
n\,\delta x_i\, + \,(N-n)\,\delta x_{i+1} &\,=\,& \Delta X\\
n\,\delta y_i\, + \,(N-n)\,\delta y_{i+1} &\,=\,& \Delta Y
\end{eqnarray}
Solving this for $N$ leads to
\begin{equation}
N=\frac{\Delta Y\, - \,r\, \Delta X}{\delta y_{i+1}\,-\,r\,\delta x_{i+1}}
\label{eq:umgebung1}
\end{equation}
where
\begin{equation}
r=\frac{\delta y_{i+1}\, -\, \delta y_{i}}{\delta x_{i+1}\, -\, \delta x_{i}}
\end{equation}
In the intervall [$0,\pi/4$] $r$ can only take the values $\infty$ (range A in Fig. \ref{fig:umgebung4}) and 
-1 (range B) which in the latter case in equation (\ref{eq:umgebung1}) 
has to be understood as limit. $r$ is the local gradient of the border of the neighborhood.\\
Since speed is distance (in cells) over number of rounds to move that distance, one has
\begin{equation}
v(\phi)=\frac{D}{N}=\frac{\delta y_{i+1}\,-\,r\,\delta x_{i+1}}{\sin{\phi}\, -\, r\,\cos{\phi}}
\end{equation}
See Fig. \ref{fig:vphi} for the speed's dependence on the direction of motion of all three complete $v=2$ neighborhoods.\\
\\
\vspace{12pt}
\subsection{The integrals for the average speeds and the deviations}
For ranges with the same gradient of the border as in {\it range A} (vertical, see Fig. \ref{fig:umgebung4}) to get the average one has to integrate:
\begin{eqnarray}
I_i^A=\int_{\phi=\phi_i}^{\phi_{i+1}} \frac{1}{\cos{\phi}} d\phi&=&\ln{\bigg(\frac{\tan(\frac{\phi_{i+1}}{2}+\frac{\pi}{4})}{\tan(\frac{\phi_i}{2}+\frac{\pi}{4})}\bigg)}\\
&=&\ln{\bigg(\frac{\sqrt{1+\tan^2{\phi_{i+1}}}+\tan{\phi_{i+1}}}{\sqrt{1+\tan^2{\phi_i}}+\tan{\phi_i}}\bigg)}
\end{eqnarray}
and for a range like {\it range B} (diagonal):
\begin{eqnarray}
I_i^B&=&\int_{\phi=\phi_i}^{\phi_{i+1}} \frac{1}{\sin{\phi}+\cos{\phi}} d\phi=\frac{1}{\sqrt{2}}\ln{\frac{\tan(\frac{\phi_{i+1} }{2}+\frac{\pi}{8})}{\tan(\frac{\phi_i}{2}+\frac{\pi}{8})}}\\
&=&\frac{1}{\sqrt{2}}\ln{\Bigg(\frac{\big(\sqrt{2(1+\tan^2{\phi_{i+1}})}-1+\tan{\phi_{i+1}}\big)\big(1+\tan{\phi_i}\big)}{\big(1+\tan{\phi_{i+1}}\big)\big(\sqrt{2(1+\tan^2{\phi_i})}-1+\tan{\phi_i}\big)}\Bigg)}
\end{eqnarray}
{\it NB: To get the average speed in that range additionally one would have to normalize the integrals.}\\
The average for the whole first octant is the sum
\begin{equation}
v_{av}=\frac{4}{\pi}\sum_i I_i^X(\phi_i,\phi_{i+1})
\end{equation}
For the deviation integrals
\begin{equation}
\int_{\phi=\phi_i}^{\phi_{i+1}} (v(\phi)-v_{av})^2 d\phi
\end{equation}
the additionally needed integrals are
\begin{equation}
\int_{\phi=\phi_i}^{\phi_{i+1}} \frac{1}{\cos{\phi}^2} d\phi=\tan{\phi_{i+1}}-\tan{\phi_i}
\end{equation}
and 
\begin{eqnarray}
\int_{\phi=\phi_i}^{\phi_{i+1}} \frac{1}{(\sin{\phi}+\cos{\phi})^2} d\phi\!\!&\!=\!&\!\!\frac{1}{2}\bigg(\tan{\Big(\phi_{i+1}-\frac{\pi}{4}\Big)}-\tan{\Big(\phi_i-\frac{\pi}{4}\Big)}\bigg)\\
\!\!&\!=\!&\!\!\frac{1}{2}\bigg(\frac{1+\tan{\phi_{i+1}}}{1-\tan{\phi_{i+1}}}-\frac{1+\tan{\phi_i}}{1-\tan{\phi_i}}\bigg)
\end{eqnarray}

\subsection{Results}
However all integrals are simple and analytic, the analytic results do not provide too much insight and in the 
following only the numerical results are given.
The following average speeds and deviations were calculated:\\
\begin{center}
\tablecaption{Average speeds and relative deviations (by angle) for all complete neighborhoods up to v=10}
\tablefirsthead{
\multicolumn{1}{c}{neighb.}& \multicolumn{1}{|c}{average}& \multicolumn{1}{|c||}{relative}&
\multicolumn{1}{|c}{neighb.}& \multicolumn{1}{|c}{average}& \multicolumn{1}{|c}{relative} \\ 
\multicolumn{1}{c}{($d_{max}^2$)}& \multicolumn{1}{|c}{speed}  & \multicolumn{1}{|c||}{deviation}&
\multicolumn{1}{|c}{($d_{max}^2$)}& \multicolumn{1}{|c}{speed}  & \multicolumn{1}{|c}{deviation}\\  \hline
}
\tablehead{
\multicolumn{6}{l}{\small \sl...continued...}\\ 
neighb. &average& relative&neighb. &average& relative\\
($d_{max}^2$) & speed & deviation&($d_{max}^2$) & speed & deviation\\   \hline
}
\tabletail{  \hline
\multicolumn{6}{r}{\small \sl...to be continued...}\\
}
\tablelasttail{}
\begin{supertabular}{c|c|c||c|c|c}
  1 & 0.79 & 0.105 & 2 & 1.12 & 0.105 \\ 
  4 & 1.59 & 0.105 &  5 & 2.11 & 0.033 \\ 
  8 & 2.24 & 0.105 &  9 & 2.52 & 0.080 \\ 
 10 & 2.98 & 0.033 & 13 & 3.28 & 0.067 \\ 
 16 & 3.47 & 0.055 & 17 & 3.82 & 0.054 \\ 
 18 & 3.91 & 0.043 & 20 & 4.22 & 0.033 \\ 
 25 & 4.57 & 0.064 & 26 & 4.85 & 0.039 \\ 
 29 & 5.11 & 0.024 & 32 & 5.17 & 0.028 \\ 
 34 & 5.40 & 0.054 & 36 & 5.52 & 0.043 \\ 
 37 & 5.75 & 0.034 & 40 & 5.97 & 0.033 \\ 
 41 & 6.13 & 0.026 & 45 & 6.33 & 0.033 \\ 
 49 & 6.43 & 0.030 & 50 & 6.67 & 0.035 \\ 
 52 & 6.86 & 0.039 & 53 & 7.05 & 0.019 \\ 
 58 & 7.22 & 0.024 & 61 & 7.35 & 0.034 \\ 
 64 & 7.44 & 0.030 & 65 & 7.77 & 0.029 \\ 
 68 & 7.94 & 0.019 & 72 & 7.98 & 0.021 \\ 
 73 & 8.13 & 0.024  &74 & 8.29 & 0.023 \\ 
 80 & 8.44 & 0.033  & 81 &  8.52 & 0.028 \\ 
 82 &  8.66 & 0.026 & 85 &  8.92 & 0.023 \\ 
 89 &  9.06 & 0.025 & 90 &  9.20 & 0.015 \\ 
 97 &  9.34 & 0.025 & 98 &  9.37 & 0.026 \\ 
100 &  9.57 & 0.030 & 101 &  9.70 & 0.025 \\ 
104 &  9.83 & 0.021 & 106 &  9.96 & 0.019 \\ 
109 & 10.09 & 0.014 & 113 & 10.18 & 0.019 \\ 
116 & 10.31 & 0.023 & 117 & 10.43 & 0.024 \\ 
\end{supertabular}
\end{center}
\setlength{\parindent}{0pt} However there are ambiguities, these information point to this choice of neighborhoods (of which only the ones for speeds 1,2,3,5 and 6 can be composed out of the corresponding number of subsequent von Neumann or Moore steps):\\
\begin{center}
\begin{tabular}{cc}
speed & neighborhood ($d_{max}^2$) \\ \hline
1 & 2 \\ 
2 & 5 \\ 
3 & 10 \\ 
4 & 18 \\ 
5 & 29 \\ 
6 & 40 \\ 
7 & 53 \\ 
8 & 72 \\ 
9 & 89 \\ 
10& 109 \\ 
\end{tabular}
\end{center}
resulting in this quarter of speed neighborhoods (an agent with maximum speed $v_m$ can reach all cells with a number $\leq v_m$):
\begin{center}
\begin{tabular}{|c|c|c|c|c|c|c|c|c|c|c|}\hline
10&10&10&10&  &  &  &  &  &  &  \\ \hline
9 &9 &9 &10&10&10&  &  &  &  &  \\ \hline
8 &8 &8 &9 &9 &9 &10&  &  &  &  \\ \hline
7 &7 &7 &8 &8 &9 &9 &10&  &  &  \\ \hline
6 &6 &6 &7 &7 &8 &8 &9 &10&  &  \\ \hline
5 &5 &5 &6 &7 &7 &8 &9 &9 &10&  \\ \hline
4 &4 &5 &5 &6 &7 &7 &8 &9 &10&  \\ \hline
3 &3 &4 &4 &5 &6 &7 &8 &9 &10&10\\ \hline
2 &2 &3 &4 &5 &5 &6 &7 &8 &9 &10\\ \hline
1 &1 &2 &3 &4 &5 &6 &7 &8 &9 &10\\ \hline
0 &1 &2 &3 &4 &5 &6 &7 &8 &9 &10\\ \hline
\end{tabular}
\end{center}
\section{A model of pedestrian motion}
As it is not in the main scope of this paper, the model which makes use of the ideas above will now be presented only in short:\\
Space becomes discretized into quadratic cells with 40 cm as length of an edge. Each cell may be occupied by at most one agent.\\
Before the beginning of the main part of the simulation, the individual parameters -- as the maximum speed -- are spread over all agents.\\
Then the agents are assigned (deterministically or randomly) to their starting position.\\
Round by round the agents repeat the following steps until all agents have left the scenario via an exit:\\
\begin{itemize}
\item All agents in parallel choose one of the cells ({\it destination}) within the neighborhood assigned to their maximum speed.\\
\item The agents sequentially try to reach their destination cell.
\end{itemize}
The rules for the selection of a destination cell are quite complex, while the rules of movement are rather simple.
Former ones are probabilistic. Most important for the decision process is the higher probability to select a cell as destination if 
it lies closer to the exit (probability $p\propto \exp(k_S (S_{max}-S))$, with coupling strength $k_S$ and distance $S$ to exit). But also herding
behavior, inertia, and the distance towards other agents as well as walls can play a role. 
In many aspects this part of the model is a higher speeds extension of the model described in Refs. \cite{Kirchner2003b,Nishinari2004}.\\
The rules of actual motion are deterministic, however the sequence in which the agents carry out their steps is chosen randomly. 
Each agent moves within a Moore neighborhood to that cell that lies closest to his destination
cell. If no cell is available that is closer to the destination cell than his current position he remains where he is. A once used cell 
remains blocked for the rest of the round (see Ref. \cite{Kluepfel2003phd}).
\section{Testing the symmetry and the discretization artifacts}
\subsection{Walking speeds and travel times of single agents}
To test the benefit of the considerations above for the equality of directions several simulations were carried out, where one agent moved
a distance of 325 (the number $< 1000$ with the most solutions of Pythagoras: $325^2=A^2+B^2$) cells into eight different directions with two different speeds. Each simulation was carried out 100 times. $k_S$ has been set to
10.0 to make the simulation nearly deterministic.
\begin{center}
\begin{tabular}{c|cccccccc}\hline
$\Delta x$    &  253&  260&  280&  300&  312&  315&  323&  325\\
$\Delta y$    &  204&  195&  165&  125&   91&   80&   36&    0\\
$\rightarrow$ angle&38.9°&36.9°&30.5°&22.6°&16.3°&14.3°&6.4°&0°    \\ \hline
$<T_{v=1}>$   &\bf 274.2 \bf &276.2&285.8&303.8&313.8&316.5&324.1&\bf 326.0 \bf\\
 $\pm St.D.$  &  3.7&  3.3&  1.8&  1.4&  0.9&  0.7&  0.2&  0.1\\
$<T_{v=5}>$   &\bf 67.5 \bf & 67.0& 66.0& \bf 64.9 \bf & 65.0& 65.2& 66.0& 66.0\\
 $\pm St.D.$  &  0.5&  0.2&  0.2&  0.4&  0.4&  0.4&  0.0&  0.0\\
\end{tabular}
\end{center}
So the overall average evacuation time for $v=1$ was 302.6 rounds $\pm$ 7.00 rounds (2.30 \%), while for $v=5$ it was 65.9 rounds $\pm$
0.30 rounds (0.46\%). This means that standarddeviation of the overall average is roughly by the same factor (5) smaller as the speed is larger.
If one wants to interpret the agents moving in the two examples with 2 m/sec for $v=1$ one round has to be interpreted as 0.2 seconds and for $v=5$
one round would be one second. Then for $v=1$ the time to move as far as 130 m would vary with the orientation of the discretization axis by more than
10 seconds ((326.0-275.2)*0.2), while for $v=5$ it would be only 2.5 seconds (67.4-64.9).
\subsection{A radially moving crowd}
\begin{figure}[htbp]
\begin{center}
\includegraphics[width=151pt]{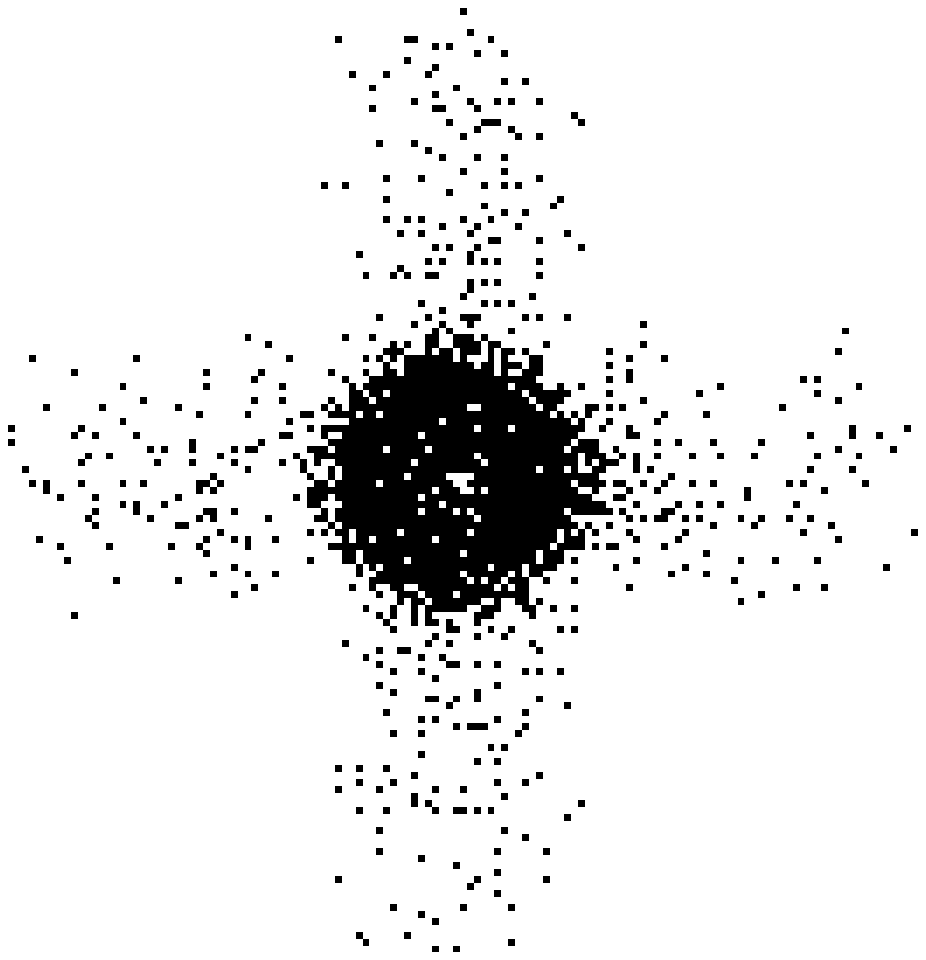}\hspace{12pt}
\includegraphics[width=151pt]{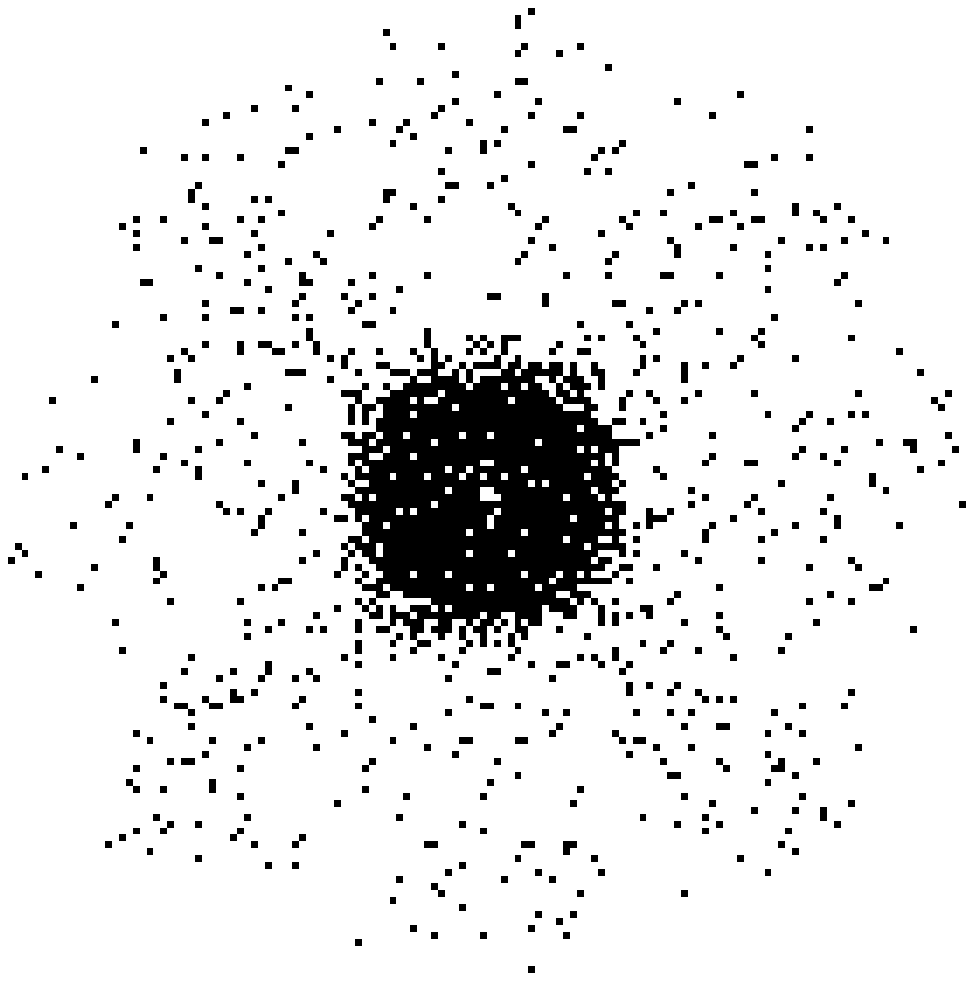}\hspace{12pt}
\caption{\it Comparison of two simulations with a crowd (black) moving to the center of a circle. The left image shows v=1 agents after 180, the right one v=5 agents after 36 rounds.}
\label{fig:shape}
\end{center}
\end{figure}

\begin{figure}[htbp]
\begin{center}
\includegraphics[width=200pt]{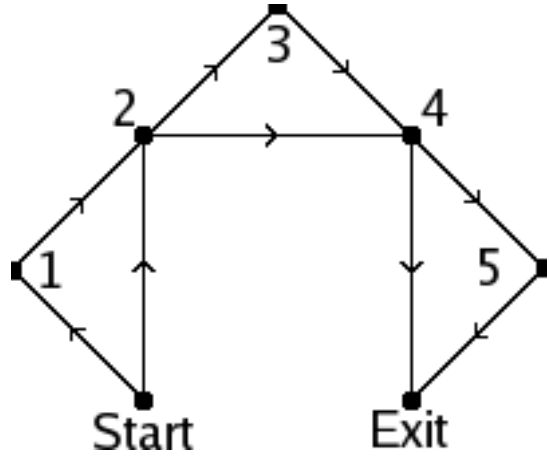}
\caption{\it Two routes: Route A contains horizontal and vertical, route B diagonal parts.}
\label{fig:quadrat}
\end{center}
\end{figure}

1948 agents were spread over 194812 cells of a circle area (radius 249 cells). With four exit cells in the center of the circle the agents started to move at once towards the center of the circle. The calculation was done twice: At first all agents had a maximum speed 1, during the second run, they had a maximum speed 5. See Fig.s \ref{fig:shape} for a comparison of how the initially rotationally symmetric spatial distribution of agents evolves with time in the two cases.
\\
As a last test we compared the simulated walking times of the two alternative routes (called A: Start $\rightarrow$ 2 $\rightarrow$ 4 $\rightarrow$ Exit and B: Start $\rightarrow$ 1 $\rightarrow$ 3 $\rightarrow$ 5 $\rightarrow$ Exit) shown in Fig. \ref{fig:quadrat}. In reality route B is $\sqrt{2}$ times as long as route A and so should the walking times be for pedestrians with identical speeds. So this is a specific comparison of motion into the two directions $0$ and $45$ degree.
\\
These are the average (ten simulations) walking times for agents with a certain speed:
\begin{center}
\begin{tabular}{c|cccc}
     &$T_A$&$T_B$&$T_B/T_A$&$T_B/(\sqrt{2}T_A)$\\ \hline
$v=1$&291.1&328.4&1.13&0.80\\
$v=2$&147.0&202.4&1.38&0.98\\
$v=3$& 98.6&155.2&1.57&1.11\\
$v=4$& 74.2&102.9&1.39&0.98\\
$v=5$& 59.4& 86.7&1.46&1.03\\
\end{tabular}
\end{center}
The deviations from 1 in the last column are due to the integer valuedness of the static floor field which leads to probabilistic path-choosing-behavior even in the deterministic limt $k_S \rightarrow \infty$. For a real valued static floor field the last column would contain only 1.00s.

\section{Summary}
In means of minimizing artifacts of discretization, we presented two criteria to identify the best neighborhoods for speeds larger one. We presented
the results of simulations which compared motion in Moore neighborhood steps with motion in steps within the best neighborhood for $v=5$. Depending
on the observable the results showed the reduction of discretization artifacts by a factor of four or even five for the latter neighborhood. This becomes specifically interesting in case of finer discretizations, where a subsequent execution of steps within Moore or von Neumann neighborhoods would lead to the same dependence of evacuation times on the orientation of the axis of discretization, yet on smaller space-scales.

\bibliographystyle{unsrt}
\bibliography{referenzen}

\end{document}